\documentclass[12 pt, amsfonts, amssymb,color]{article}

\usepackage{amsmath,amssymb,amsfonts,latexsym,float,graphics,epsfig}

\begin{document}

\renewcommand\theequation{\arabic{section}.\arabic{equation}}
\catcode`@=11 \@addtoreset{equation}{section}
\newtheorem{axiom}{Definition}[section]
\newtheorem{theorem}{Theorem}[section]
\newtheorem{axiom2}{Example}[section]
\newtheorem{lem}{Lemma}[section]
\newtheorem{claim}{Claim}[section]
\newtheorem{prop}{Proposition}[section]
\newtheorem{cor}{Corollary}[section]
\newcommand{\be}{\begin{equation}}
\newcommand{\ee}{\end{equation}}

\newcommand{\equal}{\!\!\!&=&\!\!\!}
\newcommand{\rd}{\partial}
\newcommand{\g}{\hat {\cal G}}
\newcommand{\bo}{\bigodot}
\newcommand{\res}{\mathop{\mbox{\rm res}}}
\newcommand{\diag}{\mathop{\mbox{\rm diag}}}
\newcommand{\Tr}{\mathop{\mbox{\rm Tr}}}
\newcommand{\const}{\mbox{\rm const.}\;}
\newcommand{\cA}{{\cal A}}
\newcommand{\bA}{{\bf A}}
\newcommand{\Abar}{{\bar{A}}}
\newcommand{\cAbar}{{\bar{\cA}}}
\newcommand{\bAbar}{{\bar{\bA}}}
\newcommand{\cB}{{\cal B}}
\newcommand{\bB}{{\bf B}}
\newcommand{\Bbar}{{\bar{B}}}
\newcommand{\cBbar}{{\bar{\cB}}}
\newcommand{\bBbar}{{\bar{\bB}}}
\newcommand{\bC}{{\bf C}}
\newcommand{\cbar}{{\bar{c}}}
\newcommand{\Cbar}{{\bar{C}}}
\newcommand{\Hbar}{{\bar{H}}}
\newcommand{\cL}{{\cal L}}
\newcommand{\bL}{{\bf L}}
\newcommand{\Lbar}{{\bar{L}}}
\newcommand{\cLbar}{{\bar{\cL}}}
\newcommand{\bLbar}{{\bar{\bL}}}
\newcommand{\cM}{{\cal M}}
\newcommand{\bM}{{\bf M}}
\newcommand{\Mbar}{{\bar{M}}}
\newcommand{\cMbar}{{\bar{\cM}}}
\newcommand{\bMbar}{{\bar{\bM}}}
\newcommand{\cP}{{\cal P}}
\newcommand{\cQ}{{\cal Q}}
\newcommand{\bU}{{\bf U}}
\newcommand{\bR}{{\bf R}}
\newcommand{\cW}{{\cal W}}
\newcommand{\bW}{{\bf W}}
\newcommand{\bZ}{{\bf Z}}
\newcommand{\Wbar}{{\bar{W}}}
\newcommand{\Xbar}{{\bar{X}}}
\newcommand{\cWbar}{{\bar{\cW}}}
\newcommand{\bWbar}{{\bar{\bW}}}
\newcommand{\abar}{{\bar{a}}}
\newcommand{\nbar}{{\bar{n}}}
\newcommand{\pbar}{{\bar{p}}}
\newcommand{\tbar}{{\bar{t}}}
\newcommand{\ubar}{{\bar{u}}}
\newcommand{\utilde}{\tilde{u}}
\newcommand{\vbar}{{\bar{v}}}
\newcommand{\wbar}{{\bar{w}}}
\newcommand{\phibar}{{\bar{\phi}}}
\newcommand{\Psibar}{{\bar{\Psi}}}
\newcommand{\bLambda}{{\bf \Lambda}}
\newcommand{\bDelta}{{\bf \Delta}}
\newcommand{\p}{\partial}
\newcommand{\om}{{\Omega \cal G}}
\newcommand{\ID}{{\mathbb{D}}}
\newcommand{\pr}{{\prime}}
\newcommand{\prr}{{\prime\prime}}
\newcommand{\prrr}{{\prime\prime\prime}}
\title{Reduction and Hamiltonian aspects of a model for virus-tumour interaction in oncolytic virotherapy}
\author{A Ghose Choudhury\footnote{E-mail aghosechoudhury@gmail.com}\\
Department of Physics, \\
Diamond Harbour Women's University,\\  D. H Road, Sarisha 743368, West-Bengal, India\\
\and
Partha Guha\footnote{E-mail: partha@bose.res.in}\\
SN Bose National Centre for Basic Sciences \\
JD Block, Sector III, Salt Lake \\ Kolkata 700098,  India \\
}

\date{ }

 \maketitle

\smallskip

\smallskip

\begin{abstract}
\textit{We analyse the Hamiltonian structure of a system of first-order ordinary differential equations used for modeling the interaction of an oncolytic virus with a tumour cell population. The analysis is based on the existence of a Jacobi Last Multiplier for the system  and a time dependent first integral. For suitable conditions on the model parameters this allows for the reduction of the problem to a planar system of equations for which the time dependent Hamiltonian flows are described. The geometry of the Hamiltonian flows are finally investigated using the symplectic and cosymplectic methods.}
\end{abstract}

\smallskip

\paragraph{Mathematics Classification (2010)}:34C14, 34C20.

\smallskip


\section{Introduction}
In recent years there has been a growing interest in the use of viruses for the treatment of cancer.
Oncolytic virotherapy is an emerging anti-cancer treatment modality that uses Oncolytic Viruses (OVs).
One of the salient features of the OVs is that they are either naturally occurring or genetically engineered to
selectively infect, replicate in and damage tumor cells while leaving normal cells intact.

Mathematical models have been developed for describing the interaction of tumour cells with virus
particles bioengineered to infect and destroy cancerous tissues. Naturally occurring cancer killing
viruses have shown promise in clinical trials for a number of cancer types \cite{Russell}. \\
Mathematical models have been frequently used to gain understanding of the long-term behaviour of tumour cells
under different therapies. One of the first mathematical models for oncolytic virotherapy  was deveoped by Wodraz \cite{Wodarz1, Wodarz2}. Several other researchers have been involved in developing suitable models and estimating the values of the different parameters by optimizing their model to available clinical data. The models developed by Bajzer et al \cite{Bajzer} and Titze \cite{Titze} have provided insight into the long-term behaviour of virus-tumour interaction. Based on \cite{Titze}, Jenner et al \cite{Jenner} have presented a reduced system of ordinary differential equations (ODEs) that model the interaction of an  oncolytic virus with a tumour cell population. They have numerically investigated the model dynamics focussing on a local stability analysis and bifurcations.  \\
Our motivation is to examine analytically the features of the model system of ODEs introduced in \cite{Jenner}. We will study
the Lagrangian and Hamiltonian of the reduced  virus-tumour interaction equation in oncolytic virotherapy. We obtain time
dependent Hamiltonian and explores the geometrical properties. This work demonstrates the importance of
geometrical mechanics to understand mathematical model of Oncolytic virotherapy.

\section{The model equations}
The model introduced in \cite {Jenner} is a system of three first-order ODEs which describe the interaction between
oncolytic virus and a growing tumour obeys a system of ordinary differential equation
\begin{eqnarray}\label{ME1}
\frac{dU}{dt}=\xi U-UV\\
\label{ME2}\frac{dJ}{dt}=U V - J\\
\label{ME3}\frac{dV}{dt}=-mV+J
\end{eqnarray}
Here $U, J, V$ represent in dimensionless form the uninfected tumour cell, the virus-infected tumour cell and the free virus populations respectively and  $t$ the time. $\xi$ and $m$ are dimensionless parameters of the model.
While it is acknowledged that  first-order ODEs do not provide information on spatial spread they do however provide a structure by means of which the mean-field interactions between
tumour cells and virus particles can be reasonably explored.

\subsection{First integrals and reduction to a planar system}

We begin our analysis by showing that the above system admits a time-dependent first integral.
\begin{prop}The system of equations (\ref{ME1})-(\ref{ME3}) admits a time-dependent first integral  given by
$$I=e^{mt}(U+J-(m-1)V)\;\;\;\mbox{for}\;\;\xi=-m$$
\end{prop}
{\bf Proof:}  By a direct calculation. $\Box$ \\

Introducing the following change of variables
\be\label{CoV} x=Ue^{mt}, y=Je^{mt}, z=Ve^{mt},\ee
the system (\ref{ME1})-(\ref{ME3}) reduces to  (with $\xi = -m$),
\be
\dot{x} =-xze^{-mt}\ee
\be
\dot{y}  =xze^{-mt}+(m-1)z\ee
\be
\dot{z} =y.\\
\ee
Under the above change of variables the time-dependent first integral ${\cal I}$ in new coodinates assumes a time independent form, \textit{viz}
$${\cal I} = x+y-(m-1)z.$$
As we are dealing with a system of three first-order ODEs and have succeeded in finding one first integral it follows that we can obtain another first integral provided there exists a Jacobi Last Multiplier (JLM) for the system. This is a consequence of the fact that the given a system of $n$ first-order ODEs if we can find $n-2$ first integrals and a JLM then the system may be reduced to quadrature \cite{Gor, GCG1}. The defining equation for the JLM for a non-autonomous system of first-order ODEs  given in general by
$$\dot{x}_i = f_i(x_1, ..., x_n, t)\;\;\;i=1,...,n$$
is
\be\frac{d}{dt}\log M+\sum_{i=1}^n\frac{\partial f_i}{\partial x_i}=0\ee
 In the  present case it follows that the solution for the JLM is
 \be M=\frac{e^{-(m-1)t}}{x}.\ee
Therefore on the level surface, ${\cal I}_c = c$, the above system of equations reduces to the planar system:
\begin{eqnarray} \dot{x}=-xz e^{-mt} :=f(x,z, t)\label{p1}\\
\dot{z}=c-x+(m-1)z :=g(x,z,t),
\label{p2}\end{eqnarray}
where $f$ and  $g$ are smooth explicitly time dependent real valued functions.

\subsection{Lagrangian of the reduced system}

We may
associate with the system of planar equation (\ref{p1}) and (\ref{p2}) the vector field
$${\cal X} := \frac{\partial}{\partial t} + f(x,y,t)\frac{\partial}{\partial x} +
g(x,y,t)\frac{\partial}{\partial y} $$
defined on,  ${\cal M} \times {\Bbb R}$, whose integral curves are determined
by the above system of equations. Here ${\cal M}$ denotes a real two dimensional manifold with local coordinates $x$ and $y$.
It is interesting to note that the planar system defined on the level curves, ${\cal I}=c$, by (\ref{p1})-(\ref{p2}) admits a Lagrangian description. By eliminating the variable $z$ one arrives at the following second-order ODE in the variable $x$, namely
\be\label{L1} \ddot{x}-\frac{\dot{x}^2}{x}+\dot{x}+x(c-x)e^{-mt}=0.\ee
The JLM for  an equation of the form, $\ddot{x}=F(x, \dot{x}, t)$, is defined as a solution of the following equation
$$\frac{d\log \tilde{M}}{dt}+\frac{\partial F}{\partial \dot{x}}=0.$$
In the present case this yields
\be \tilde{M}=\frac{e^t}{x^2}.\ee
Note that as $\tilde{M}=\partial^2L/\partial\dot{x}^2$ it follows that a Lagrangian for the reduced system is given by
\be L(x, \dot{x}, t)=\frac{e^t\dot{x}^2}{2x^2}-e^{(m-1)t}[c\log x-x].\ee

\smallskip

The generalized variational problem proposed by Herglotz in 1930 \cite{Herglotz}, deals with an initial value problem
$$ \dot{u}(t) = L(t,x(t),\dot{x}(t),u(t)), \qquad t \in [a,b] $$
with $u(a) = \gamma$, $\gamma \in {\Bbb R}$,
consists in determining trajectories $x$ subject
to some initial condition $x(a) = \alpha$ that extremize (minimize or maximize) the value $u(b)$,
where $L \in C^1\big([a,b] \times {\Bbb R}^{2n +1}, {\Bbb R} \big)$.

Herglotz proved that a necessary optimality condition for a pair $(x(·), z(·))$ to be an extremizer of
the generalized variational problem \cite{Herglotz,Herglotz1,Kuwabara}
\be \frac{d}{dt}\big(\frac{\partial L}{\partial \dot{x}}\big) - \frac{\partial L}{\partial x} =
\frac{\partial L}{\partial \dot{x}}\frac{\partial L}{\partial u}. \ee
This equation is known as the generalized Euler-Lagrange equation. Note that
for the classical problem of the calculus of variations one has $\frac{\partial L}{\partial u} = 0$.

\smallskip

We obtain the equation of motion via generalized Euler-Lagrange equation setting $ u = t$.
If we choose to eliminate $x$ in favour of $z$ then the corresponding second-order ODE for $z$ is just
an equation of the Li\'{e}nard type, namely
\be \ddot{z}-(ze^{-mt}-(m-1))\dot{z}-(cz-(m-1)z^2)e^{-mt}=0.\ee

\subsection{Hamiltonian aspects}

As the JLM is explicitly time dependent we next follow the procedure outlined in \cite{Torres, GCG2} to obtain the Hamiltonian structure of the resulting planar system of ODEs (\ref{p1})-(\ref{p2}).
This requires us to find functions $\psi$ and $\phi$ such that
\be\label{Exact}M((f-\psi) dz-(g-\phi)dx)=dH+\theta dt,\ee
where $H$ represents the Hamiltonian of the system and $\theta$ is some real valued function. The condition for exactness then translates to the requirement
$$\partial_x(M(f-\psi))-\partial_z(M(g-\phi))=0.$$
On substituting the expressions for $f$ and $g$ from the above planar system we find that this equality is satisfied by the following choices of the functions $\psi$ and $\phi$ namely:
$$\psi=x, \;\;\phi=(m-1)z$$
Using these expressions it follows from (\ref{Exact}) that
\be\label{Ham} H=e^{-(m-1)t}(x-z-c\log x)-e^{-(2m-1)t}\frac{z^2}{2},\ee
while $$\theta=[(m-1)e^{-(m-1)t}(x-z-c\log x)-(2m-1)e^{-(2m-1)t}\frac{z^2}{2}].$$
The canonical coordinates are then identified from the relation
\begin{align}dQ\wedge dP &=M(dx-\psi dt)\wedge (dz-\phi dt),\nonumber\\
& =\frac{e^{-(m-1)t}}{u}(dx-xdt)\wedge (dz-(m-1) z dt),\nonumber\\
&=d(\log x -t)\wedge d(ze^{-(m-1) t}),\nonumber\end{align}
so that we have finally
\be Q=\log x-t,\;\;P=ze^{-(m-1) t}.\ee
In terms of the canonical variables the Hamiltonian (\ref{Ham}), written as $\tilde H$, may be expressed in the form
\be\label{HamCan}{\tilde H} = e^{Q-(m-2)t}-P-c(Q+t)e^{-(m-1)t}-\frac{P^2}{2}e^{-t}.\ee
The Hamiltons equations are therefore given by
\be\label{ham1}\dot{Q}=\frac{\partial \tilde H}{\partial P}=-1-Pe^{-t},\ee
\be\label{ham2}\dot{P}=-\frac{\partial \tilde H}{\partial Q}=-e^{Q-(m-2)t}+ce^{-(m-1)t}.\ee
In the next section we investigate the geometry of the Hamiltonian flow.

\subsection{Poincar\'e-Cartan form and time-dependent Hamiltonian flow}

The distinguished role of the time $t$ is not desirable in the
general case of non-autonomous Hamiltonian systems. We shall therefore
introduce an evolution parameter $s$ that parameterizes the
 time evolution of the system. In the extended formalism the time $t$ is
treated as an ordinary canonical function $t(s) \equiv x^{0}(s)$
of a evolution parameter $s$. Furthermore we conceive of a `new' momentum
coordinate $p_0 (s)$ in conjunction with the time $t$  as an additional
pair of canonically conjugate coordinates \cite{Kuwabara,Massa}.
The extended
Hamiltonian ${\cal H}(q^0,p_0,q^i,p_i)$ is then  defined as a
differentiable function on the cotangent bundle $T^{\ast}Q =
T^{\ast}({\Bbb R} \times M)$  with$\frac{\partial {\cal H}}{\partial s} =
0$. It is given by ${\cal H}(q^0,p_0,q^i,p_i) = H(q^i,p_i,q^0) +
p_0$, where $q^0$ and $p_0$ are conjugate variables and $p_0 = -H
+ K$, with $K$ being a constant.

The extended phase space admits a
Liouville form \be\label{4.1} {\cal \theta}_{\cal H} = p_0dt +
p_idq^i \ee and the Hamiltonian flow is completely determined by
the conditions:
$$
<{\Bbb X}_{\cal H}, dt > = 1\;\;\; \mbox{and} \qquad {\Bbb
X}_{\cal H} \lrcorner d{\cal \theta}_{\cal H} = 0,
$$
where ${\Bbb X}_{\cal H}$ is the Hamiltonian vector field. It is defined by
\be
{\Bbb X} = \frac{\partial {\cal H}}{\partial x^i}\frac{\partial }{\partial p_i}
- \frac{\partial {\cal H}}{\partial p_i}\frac{\partial }{\partial x^i}
+ \frac{\partial {\cal H}}{\partial t}\frac{\partial }{\partial p_0}
- \frac{\partial {\cal H}}{\partial p_0}\frac{\partial }{\partial t}.
\ee

The symplectic $2$-form $\Omega = d{\cal \theta}_{\cal H}$ makes the
extended space a $(2n+2)$-dimensional symplectic manifold endowed
with a Poisson bracket \be \{f,g\}_e = \frac{\partial f}{\partial
t}\frac{\partial g}{\partial p_0} + \frac{\partial f}{\partial
q^i}\frac{\partial g}{\partial p_i} - \frac{\partial f}{\partial
p_i}\frac{\partial g}{\partial q^i} - \frac{\partial f}{\partial
p_0}\frac{\partial g}{\partial t}. \ee Considering ${\cal H} = p_0
+ H$,  we obtain \be \{f,{\cal H} \}_e = \frac{\partial
f}{\partial t} + \{f,H\} = {\Bbb X}_{\cal H}(f), \ee where the
time-dependent Hamiltonian vector field is given by \be\label{vf}
{\Bbb X}_{\cal H} = \frac{\partial}{\partial t} + \{\cdot , {H}
\}= \frac{\partial}{\partial t} + \frac{\partial H}{\partial
p_i}\frac{\partial}{\partial q^i} - \frac{\partial H}{\partial
q^i}\frac{\partial}{\partial p_i}. \ee


\subsubsection{ Applications to the reduced  virus-tumour interaction equation}

In this section we apply the geometry of the time-dependent Hamiltonian system to the reduced
virus-tumour interaction planar system.
Let
\be \omega = dP \wedge dQ + dP_0 \wedge dt = dP \wedge dQ - dH \wedge dt \ee  be the symplectic form on the extended phase space,
where $P$ and $Q$ are the canonical coordinates of the reduced system and $P_0 = -H$.
The corresponding time-dependent Hamiltonian vector field corresponding is
given by
\be
X_H = \frac{\partial H}{\partial P}\frac{\partial }{\partial Q} -
\frac{\partial H}{\partial Q}\frac{\partial }{\partial P} +
\frac{\partial H}{\partial t}\frac{\partial }{\partial P_0} + \frac{\partial }{\partial t},
\ee
where $P_0 = - H$ with $H =  e^{Q-(m-2)t}-P-c(Q+t)e^{-(m-1)t}-\frac{P^2}{2}e^{-t}$.
\be
dH = \big(dQ - (m-2)dt\big)e^{Q - (m-2)t} - dP - c dQ e^{-(m-1)t} + c(Q+t)(m-1)dte^{-(m-1)t} -
PdPe^{-t} + \frac{P^2}{2} dt.
\ee
Using $\frac{\partial H}{\partial P}$, $\frac{\partial H}{\partial Q}$ from (\ref{ham1} )and (\ref{ham2} ) with
\be
\frac{\partial H}{\partial t} = -(m-2)t)e^{Q - (m-2)t} + c(Q+t)(m-1)e^{-(m-1)t} + \frac{P^2}{2}e^{-t}.
\ee
We obtain the following result.

\begin{claim}
The dynamical flow of the
system is expressed in the form of the time-dependent  Hamiltonian vector field, known as
the Hamiltonian flow, completely determined by the conditions
\be i_{X_H}\omega = -dH, \qquad i_{X_H}dt = 1. \ee
\end{claim}

\bigskip

\noindent
The symplectic form in the canonical coordinates is connected to the ``old'' coordinates via the Jacobi last multipler
in the following way.
The $dP \wedge dQ$  in terms of old coordinate can be expressed as
$$ dP \wedge dQ = M \big(dz \wedge dx + (m-1) z dx \wedge dt - x dz \wedge dt \big)
= dM \wedge dK - me^{-(m-1)t}dz \wedge dt, $$ where $K = xz$,
with  $$ dH \wedge dt = d{\tilde H} \wedge dt. $$
Thus it is clear that the symplectic form with respect to old coordinates $(x,z)$ yields
non-canonical structure, in other words this yields non-canonical Poisson bracket.

\subsection{Hamiltonian Geometric description via cosymplectic method}

A cosymplectic manifold \cite{Cap,CdLM1,CdLM2,EG} is a triple $(M,\eta, \omega)$ consisting of
a smooth $(2n+1)-$ dimensional manifold
$M$ with a closed $1$-form $\eta$ and a closed $2$-form $\omega$, i.e., $d\eta = d\omega = 0$,
such that $\eta \wedge \omega^n \neq 0 $. The Reeb field $\xi$ is
uniquely determined by $\eta(\xi) = 1$ and $i_{\xi}\omega = 0.$

Let $(M, \eta, \omega)$ be a cosymplectic manifold. Let $\phi : M \to M$ be a diffeomorphism.
Then $\phi$ is a weak cosymplectomorphism if $\phi^{\ast}\eta = \eta$ and there exists a function
$H_{\phi} \in C^{\infty}(M )$ such that $\phi^{\ast}\omega = \omega - dH_{\phi} \wedge \eta$
$\phi$ satisfies cosymplectomorphism when $H_{\phi} = 0$, i.e., $\phi^{\ast}\eta = \eta$ and
$\phi^{\ast}\omega = \omega$. Hence it respects
the Reeb field and the characteristic foliation.

\bigskip

Let $C^{\infty}(M)$ be the ring of differentiable functions on $M$, $\mathfrak{X}(M)$ and $\Omega(M)$ the
$C^{\infty}(M)$-modules of differentiable vector fields and $1$-forms of $M$, respectively.
The bundle homomorphism yields an isomorphism of $C^{\infty}(M)$-modules
$\chi : \mathfrak{X}(M) \to \Omega(M)$ defined by
\be\label{cosym}
X \in \mathfrak{X}(M) \mapsto \chi(X) = i_X\omega + \eta(X)\eta.
\ee
The Reeb vector field $\xi$ is given by $\xi = \chi^{-1}(\eta)$ and it is characterized by the identities
$i_{\xi}\omega = 0$, $\eta(X) =1$.

\bigskip

Let $(M, \eta, \omega)$ be a cosymplectic manifold, let $\xi$  denote the Reeb field and
let $X \in \mathfrak{X}(M)$ be a vector field, then $X$ is said to be weakly Hamiltonian
if $\eta(X) = 0$ and if there exists $f \in C^{\infty}(M)$ such that $i_X \omega = df - \xi(f )\eta$.
Let $H : M \to {\mathbb R}$ be a Hamiltonian function on $M$, then there exist a unique Hamiltonian
vector field $X_H$ on $M$ such that
$$\chi(X_H) = dH - \xi(H)\eta + \eta, \,\,\,\,\,\, \hbox{ where } \,\,\,\,\,\,  i_{X_H}\omega = dH - \xi(H) \eta,
\,\,\,\,\,\,  \eta(X_H) = 1.
$$
The gradient of $H$ is defined by
\be
\chi(\hbox{grad}(H) = dH,
\ee
which yields
\be \label{gradH}
\hbox{grad}(H) = \frac{\partial H}{\partial p_i} \frac{\partial }{\partial q^i} - \frac{\partial H}{\partial q^i}
\frac{\partial }{\partial p_i} + \frac{\partial H}{\partial z} \frac{\partial }{\partial z}.
\ee

The Hamiltonian vector field thus given by
\be X_H = \hbox{grad}H) - \xi(H)\xi, \ee
where $\xi$ is the Reeb vector field.
We obtain the local expression of the evolution vector field from equation (\ref{gradH})
\be
{\Bbb E}_H = \frac{\partial H}{\partial p_i} \frac{\partial }{\partial q^i} - \frac{\partial H}{\partial q^i}
\frac{\partial }{\partial p_i} + \frac{\partial }{\partial z}.
\ee
The evolution vector field ${\Bbb E}_H$ is related to Hamiltonian vector field via
\be
{\Bbb E}_H = X_{H} + \frac{\partial}{\partial t}.
\ee
Therefore, an integral curve $(q^i(t), p_i (t), z(t))$ satisfies the time-dependent Hamiltonian equations
$$ \dot{q}^i = \frac{\partial H}{\partial p_i}, \qquad \dot{p}_i = -\frac{\partial H}{\partial q^i}, \qquad
\dot{z} = 1, $$
where $\cdot$ stands for derivative with respect to $t$.

\subsubsection{Cosymplectic framework for the reduced  virus-tumour interaction equation}

In our example, the Darboux coordinates $(Q,P,t)$ are the local coordinates on the cosymplectic manifold such that
$$ \omega = dP \wedge dQ, \qquad \eta = dt, $$
and the Reeb vector field $\xi = \frac{\partial}{\partial t}$. Then the gradient of
$$ H = e^{Q-(m-2)t}-P-c(Q+t)e^{-(m-1)t}-\frac{P^2}{2}e^{-t} $$ is given by
$$
\hbox{grad}H =     \Big(-1-Pe^{-t}\Big)\frac{\partial}{\partial Q} -
\Big(e^{Q-(m-2)t}- ce^{-(m-1)t}\Big)\frac{\partial}{\partial P}  $$
$$+ \Big(-(m-2)t)e^{Q - (m-2)t} + c(Q+t)(m-1)e^{-(m-1)t} + \frac{P^2}{2}e^{-t} \Big) \frac{\partial}{\partial t}.
$$
It is clear from the definition
$\chi (\hbox{ grad }H) \longmapsto i_{\hbox{ grad}H}\omega + \frac{\partial H}{\partial t}dt = dH$,
that the contraction of $\omega$ with respect to $\hbox{ grad}H$ yields
$$ i_{\hbox{ grad}H}\omega =  dH - \frac{\partial H}{\partial t}dt = -dH + \xi(H)\eta, \qquad i_{E_H}\eta  = 1. $$
Please note that our sign is opposite to the conventional one because we have defined $\omega =  dP \wedge dQ$ instead of $dQ \wedge dP$.

\section{Summary}
In this article we have considered a model for virus-tumour
interaction in oncolytic virotherapy expressed in the form of a system of three ODEs.  In our analysis of the system we have shown the existence of a time dependent first integral for the system and also a Jacobi Last Multiplier. The existence of these two ingredients allow us to reduce the model to a planar system on the level curves. The resulting planar system is shown to admit a Hamiltonian, albeit of a time dependent variety, and one can construct canonical coordinates.
It appears that the non-existence of a time independent first integral for the original model equations prevents us from constructing
the standard Poisson structure of the system. The explicit time dependence is encompassed into the Hamiltonian framework by defining
an extended Hamiltonian formalism and explicitly demonstrating the geometric structure using Poincar\'e-Cartan two form.
This reduced time-dependent planar system is also studied in the framework of cosymplectic geometry.
Our present study compliments the investigations carried out in \cite{Jenner} revealing the rich analytical
and geometrical aspects of the model.

\section*{Acknowledgements}

PG is grateful to Professors Manuel de Leon and  Ogul Esen for various discussions and valuable remarks.
In particular, we appreciate the supportive comments and careful reading of our manuscript from Ogul Esen.

\end{document}